\documentclass[journal=jacsat,manuscript=article]{achemso}

\usepackage{chemformula} 
\usepackage[T1]{fontenc} 

\mciteErrorOnUnknownfalse
\author{Ulrich B\"ottger}
\email{boettger@iwe.rwth-aachen.de}
\phone{+49 (0)241 8027823}
\fax{+49 (0)241 8022300}
\affiliation[RWTH]
{Institut f\"ur Werkstoffe der Elektrotechnik (IWE 2), RWTH Aachen University, 52056~Aachen, Germany}
\author{Moritz von Witzleben}
\affiliation[RWTH]
{Institut f\"ur Werkstoffe der Elektrotechnik (IWE 2), RWTH Aachen University, 52056~Aachen, Germany}
\author{Viktor Havel}
\affiliation[RWTH]
{Institut f\"ur Werkstoffe der Elektrotechnik (IWE 2), RWTH Aachen University, 52056~Aachen, Germany}
\author{Karsten Fleck}
\affiliation[RWTH]
{Institut f\"ur Werkstoffe der Elektrotechnik (IWE 2), RWTH Aachen University, 52056~Aachen, Germany}
\author{Vikas Rana}
\affiliation[FZJ]
{Peter Gr\"unberg Institut (PGI-7), Forschungszentrum J\"ulich, 52425~J\"ulich, Germany}
\author{Rainer Waser}
\affiliation[RWTH]
{Institut f\"ur Werkstoffe der Elektrotechnik (IWE 2), RWTH Aachen University, 52056~Aachen, Germany}
\alsoaffiliation[FZJ]
{Peter Gr\"unberg Institut (PGI-7), Forschungszentrum J\"ulich, 52425~J\"ulich, Germany}
\author{Stephan Menzel}
\affiliation[FZJ]
{Peter Gr\"unberg Institut (PGI-7), Forschungszentrum J\"ulich, 52425~J\"ulich, Germany}

\title[An \textsf{achemso} demo]
  {Picosecond Multilevel Resistive Switching in Tantalum Oxide Thin Films}

\abbreviations{IR,NMR,UV}

\begin{document}


\begin{abstract}
The increasing demand for high-density data storage leads to an increasing interest in novel memory concepts with high scalability and the opportunity of storing multiple bits in one cell. A promising candidate is the redox-based resistive switch repositing the information in form of different resistance states. For reliable programming, the underlying physical parameters need to be understood. We reveal that the programmable resistance states are linked to internal series resistances and the fundamental nonlinear switching kinetics. The switching kinetics of Ta$_2$O$_5$-based cells was investigated in a wide range over 15 orders of magnitude from 250\,ps to 10$^5$\,s. We found strong evidence for a switching speed of 10\,ps which is consistent with analog electronic circuit simulations. On all time scales, multi-bit data storage capabilities were demonstrated. The elucidated link between fundamental material properties and multi-bit data storage paves the way for designing resistive switches for memory and neuromorphic applications.
\end{abstract}

\section{Introduction}
The class of redox-based resistive switching devices (ReRAM) based on the valence change mechanism (VCM) is a potential type for future non-volatile memory \cite{Waser2009004,Wong2012002}, and computation-in-memory applications \cite{Ielmini2018001,Yang2013001,Prezioso2015001,Siemon2015002,Borghetti2010001}. A typical VCM cell consists of a resistively switching oxide layer sandwiched between a high work function metal electrode such as Pt and a low work function metal, e.\,g.\ Ta. Among the numereous resistively switching oxides, \ch{Ta2O5} is a promising material in terms of endurance \cite{Lee2011023}, scalability \cite{Hayakawa2015001}, switching speed \cite{Torrezan2011001}, and multi-level switching capability \cite{Kim2016006}. Before the VCM cell can be switched repetitively between a high resistive state (HRS) and a low resistive state (LRS), an electroforming step is required. For this, a voltage is applied to the cell and the oxide thin film is locally reduced by extraction of oxygen resulting in a highly-conducting, oxygen-deficient filamentary region \cite{Waser2009004,Yang2009002}. 

The resistive switching effect has been attributed to a movement of mobile donors such as oxygen vacancies or cation interstitials, and a subsequent change in the filament composition leading to a valence change in the cation sublattice \cite{Waser2009004,Yang2008001,Waser2012001,Miao2012001}. As the switching mechanism is dominated by the drift of ions, the switching operation is inherently bipolar. One voltage polarity is needed to set the cell from HRS to LRS, whereas the opposite voltage polarity is required to reset the device from LRS to HRS. Typically, an abrupt set transition is observed, whereas the reset transition is gradual \cite{Fleck2016002,Marchewka2015002,Yu2011004}. The capability of multi-level operation, i.\,e.\ storing multiple bits per cell, enhances the storage density \cite{Lee2012007,Graves2017001,Kim2017003}. In that case, the programming process is stopped at a specific intermediate resistive state (IRS) and is controlled either by the applied voltage during the gradual reset of the cell \cite{Marchewka2015002,Bai2014001,Lee2012007} or by a current compliance during the set operation \cite{Li2019001,Bai2014001}. For neuromorphic applications, the feature of multi-level switching is essential \cite{Kim2012003,Burr2017001}.

In order to meet the needs for future non-volatile memories, the so-called voltage-time-dilemma has to be overcome \cite{Waser2009004}. This corresponds to an extremely nonlinear switching kinetics of the ReRAM cell charaterized by a low-voltage read-out operation over a long period up to ten years and a fast write process in the nano-second regime or below by applying a voltage that is about ten times higher than the read voltage. While several groups have studied the switching kinetics of ReRAMs in certain limited ranges as compiled in \cite{Menzel2015001}, an investigation over the complete dynamic range has not been demonstrated yet. To cover the full time-domain, the measurements have to be extended to the sub-nanosecond regime, too. Resistive switching in the sub-nanosecond regime has been qualitatively demonstrated for VCM cells based on HfO$_2$ \cite{Lee2010011}, Ta$_2$O$_5$ \cite{Torrezan2011001}, SiO$_2$ \cite{choi2013002}, and AlN \cite{Choi2016002}. The switching event, however, could not be resolved in these studies and the reproducibility of the switching on a single cell was rather low. 

Here, we present a comprehensive study of the set switching kinetics of Ta$_2$O$_5$-based VCM cells from 250\,ps to up to 10$^5$\,s by the means of an optimized coplanar waveguide (CPW) device structure and the use of multiple measurement setups. This enables us to resolve the switching time over 15~orders of magnitude at the same VCM cell.

Furthermore, we demonstrate highly reproducible multi-level programming performed by varying the amplitude and length of the pulse. The data analysis reveals that the programmed LRS is linked to the inherent nonlinear VCM switching kinetics and an internal series resistance. Based on this finding, we discuss design rules for optimizing the multi-level programming capability of VCM cells integrated with a passive selector.

\section{Results and discussion}

\subsection{Effects of series resistances}
\label{series_res} 

\begin{figure}
\includegraphics[width=\textwidth]{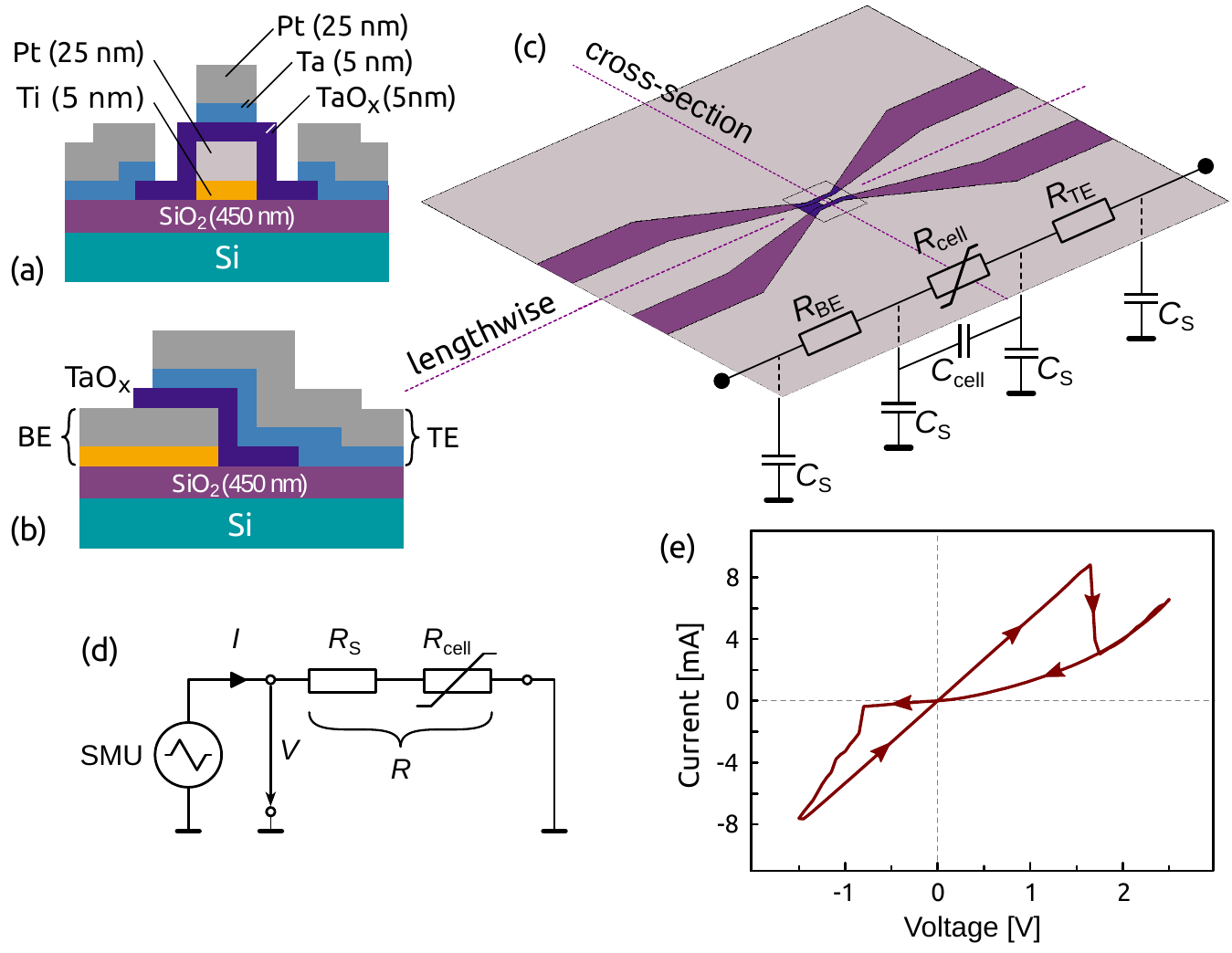}
\caption{\label{fig_1_scheme} Schematic overview of the Ta$_2$O$_5$ ReRAM cell showing (a) the cross-section, (b) the lengthwise arrangement, and (c) the integrated structure with overlapping electrodes including the corresponding $RC$ network. (d) Equivalent circuit diagram with the cell resistance $R_{\rm
cell}$ and the serial resistance $R_{\rm S} = R_{\rm{BE}} + R_{\rm TE}$ for quasi-static characterization, and (e) $I(V)$-sweep characteristic of the CPW device. A gradual set is observed between $-0.9\,$\,V$<V< -1.6$\,V, an abrupt reset appears at $V\approx+1.6$\,V.}
\end{figure}

The schemes in Figure~\ref{fig_1_scheme}\,a-d show the investigated, tapered CPW structures with Ta$_2$O$_5$ ReRAM cells optimized in terms of high frequency impedance matching. Cells with area sizes of $300$\,$\mu$m$^2$ and $600$\,$\mu$m$^2$ were measured. 
The fabrication process of the layer stack is identical with that of our previous work \cite{Kim2016007}.  
The equivalent circuit of the entire device includes the variable resistance of the Ta$_2$O$_5$ layer $R_{\rm{cell}}$ and the series resistance $R_{\rm S}$ combining the contributions of electrodes, electrical lines, and the contacts of the bottom and top electrode path. By electrical characterization, only the total device resistance $R$ of the entire device is measurable: $ R= R_{\rm S} + R_{\rm cell}$ (Figure~\ref{fig_1_scheme}\,d).

The existence of $R_{\rm S}$ leads to the fact that in a quasi-static $I(V)$-curve the characteristic abrupt set transistion of VCM devices appears as a gradual transition. This behavior is also observed for the CPW devices under test (DUT), see Figure~\ref{fig_1_scheme}\,e. The switching starts at a specific negative voltage. When the cell resistance decreases during set operation, it approaches the range of the series resistance. The applied voltage $V = V_{\rm{S}}+V_{\rm{cell}}$ will be redistributed between $R_{\rm{cell}}$ and $R_{\rm{S}}$. In consequence, the cell voltage $V_{\rm{cell}}$, and therefore, the driving force for resistance reduction decreases until the process finally grinds to a halt in the timeframe of the experiment at a defined voltage $V_{\rm{min}}$ \cite{Wouters2012001, Ielmini2012001, Menzel2013003}. The gradual set transition does not necessarily have to originate from internal series resistances, it may also be caused by external series resistances \cite{Wouters2012001,Hardtdegen2018001,Kim2016004}. It is most essential that $R_{\rm S}$ is independent of the applied voltage. 

In a similar manner the reset operation of the ReRAM cell is modified by the series resistance, see Figure~\ref{fig_1_scheme}\,e. In case of the application of a positive voltage to the device in the LRS, the series resistance is dominant and the applied voltage mainly drops over $R_{\rm S}$. As soon as the cell resistance increases during the reset operation, the ratio of the voltage divider changes and causes a positive feedback, i.\,e.\ the cell voltage increases. The reset process speeds up and the transition becomes abrupt. The series resistance, hence, masks the intrinsic abrupt set and gradual reset behavior of the VCM cell and turns it into a gradual set and an abrupt reset process \cite{Hardtdegen2018001}.

\subsection{Ultra-fast multilevel switching}

\begin{figure}
\includegraphics[width=0.75\textwidth]{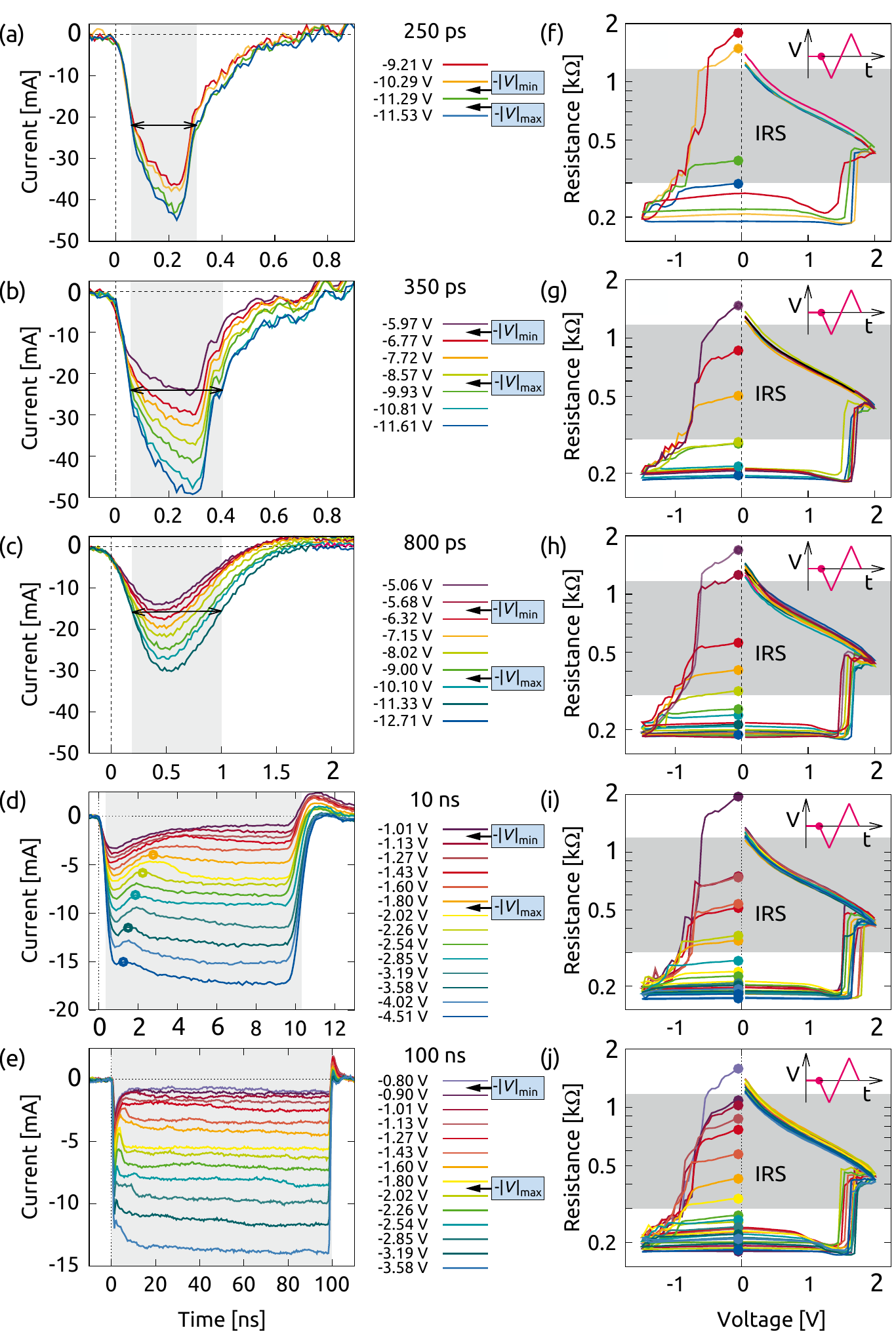}
\caption{\label{ps} (a-e) Transient pulse measurements of a $20 \times$30\,$\mu$m$^2$ Ta$_2$O$_5$ CPW device stack in the picosecond and nanosecond time range for a series of set pulses with variable pulse amplitude $|V_p|$ and (f-j) the corresponding subsequent $R(V)$-sweep measurements on the same device after applying the write (set) pulses. The $R(V)$ curves start at the dot at $-0.05$\,V either in the HRS ($R > 1.2$\,k$\Omega$) or IRS (grayed-out area) or LRS ($R < 300$\,$\Omega$) and end in the HRS for all write pulses. The multi-level operation is obvious.}
\end{figure}

A series of ultra-short set pulses with lengths between 250\,ps and 100\,ns and amplitudes up to 12.7\,V were applied to the CPW devices in the HRS. The transients in Figure~\ref{ps}\,a-e show the corresponding waveforms of the current through a cell with $A=600$\,$\mu$m$^2$. For the 10\,ns and 100\,ns pulses, the switching events for each amplitude are clearly identified and are exemplarily illustrated by the marked inflection points in Figure~\ref{ps}\,d. The general trend shows faster switching for increasing pulse amplitudes. No inflection point, i.\,e.\ no switching event, is observed as long as the absolute value $|V_p|$ is below a minimum voltage $|V|_{\rm{min}}$. Overshoots by charging and discharging the cell capacitance predominantly determine the transient currents over the full time range. 

The response on picosecond pulses is without any signature of a possibly happened switching event (Figure~\ref{ps}\,a-c). This originates from the large device capacitance, e.\,g.\ $C_{\rm{dev}}\approx 20$\,pF at $A=600$\,$\mu$m$^2$, whose charging affects the signal waveform and makes the signal changes of higher bandwidth undetectable. A similar behavior was also found by Torrezan et al.\ \cite{Torrezan2011001}. Nevertheless, even in case of 250\,ps pulses a ``complete'' switching from the HRS into the LRS was clearly demonstrated. 

The verification of the resistance reduction after pulsing was carried out by subsequent $I(V)$-sweeps with a linear rate of 0.1\,V/s, which start with the same negative polarity as the set pulse. The resulting $R(V)$ behavior is illustrated in Figure~\ref{ps}\,f-j. For low pulse amplitudes $|V_{p}| \le |V|_{\rm{min}}$, the $I(V)$-sweep of the ReRAM cell starts in the HRS ($R > 1.2$\,k$\Omega$) because the stimulus of the prior fast set pulse was not sufficiently strong enough to induce the switching process. In the range $|V|_{\rm{min}} < |V_p| < |V|_{\rm{max}}$ the cell is switched partially to an intermediate state, whose resistances monotonically decrease with increasing pulse amplitudes $|V_p|$. The cell is switched fully to the LRS defined here as $R < 300$\,$\Omega$, since $|V_p| \ge |V|_{\rm{max}}$. Further voltage enhancement will result in no or little resistance decrease. The voltage limits for the different pulse lengths can be estimated from Figure~\ref{ps}.

The programming of different resistance states by amplitude modulation was so far only observed for pulse lengths of 100\,ns or longer, e.\,g.\ \cite{Stathopoulos2017001}. Here, we could demonstrate for the first time the multilevel set capability even with picosecond pulses. This behavior was confirmed for different cells at various sizes.

\subsection{Extended time domain measurements}

\begin{figure}
\includegraphics[width=\textwidth]{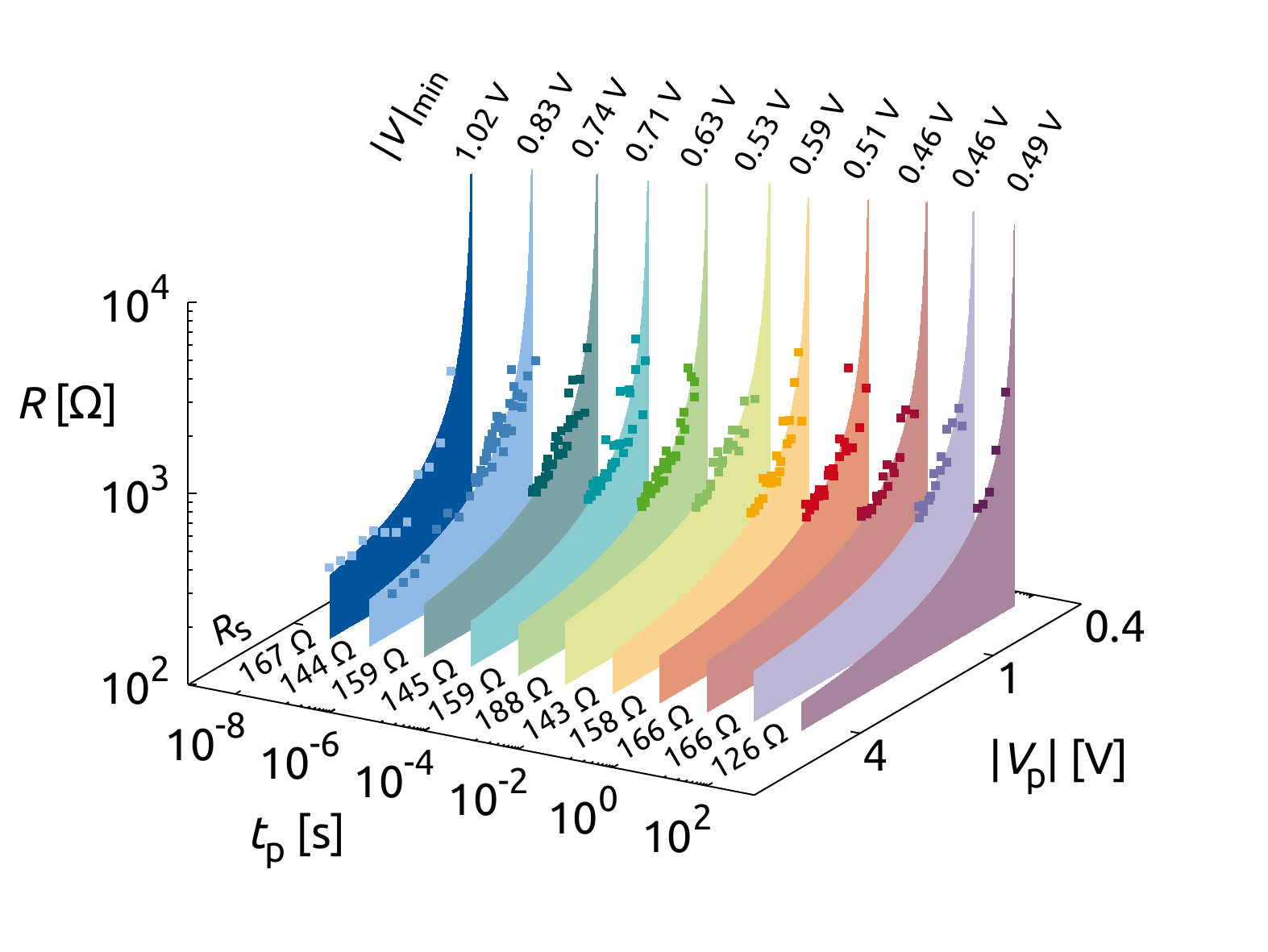}
\caption{\label{fig_3d} 3D point plot of the measured data ($R,V_p$) of a $15 \times$20\,$\mu$m$^2$ Ta$_2$O$_5$ CPW cell for a given pulse width $t_p$, and fitted behavior of the programmed resistance state $R$ using the fit parameters $|V|_{\rm min}$ and $R_{\rm S}$ pursuant to Eq.\,(\ref{Prog_R}).}
\end{figure}

The expansion of the investigation to pulse lengths up to 10$^5$\,s reveals that programming the LRS or one of the IRS by controlling the pulse amplitude is possible on all time scales. In Figure~\ref{fig_3d} the programmed resistance $R$ is plotted versus the absolute value of the applied pulse voltage $|V_p|$ for different pulse widths $t_p$. Two trends can be observed: (i) $R$ appears to be inversely proportional to $|V_p|$ for all $t_p$, and (ii) the programmed resistance becomes lower for longer pulse widths. 

As already mentioned, the series resistance reduces the voltage drop over the memory cell during set operation and leads to a certain minimum voltage $|V|_{\rm min}$, at which the driving force for further resistance reduction of $R_{\rm{cell}}$ becomes practically zero. The fact that $R$ depends on $t_p$ implies that $|V|_{\rm min}$ depends on $t_p$, too. Thus, $|V|_{\rm min}$ should be linked somehow to the switching kinetics of the device. As long as the voltage does not exceed $|V|_{\rm min}$, the cell stays in the HRS and the maximum current during set operation for a given $t_p$ can be described by Kirchhoff`s current law:
\begin{equation}
\label{imax}
|I|_{\rm{max}} = \displaystyle \frac{|V(t_{p})|_{\rm min}}{R_{\rm cell}}= \displaystyle \frac{|V_p|}{R}= \displaystyle \frac{|V_p|-|V(t_{p})|_{\rm min}}{R_{\rm S}}.
\end{equation}

\noindent Reformulating Eq.\,(\ref{imax}) provides an expression for the programmed resistance state
\begin{equation} \label{Prog_R}
R = \displaystyle \frac{|V_p|}{|V_p|-|V(t_{p})|_{\rm min}} \, {R_{\rm S}}= \displaystyle \frac{R_{\rm S}}{1 - 
|V(t_{p})|_{\rm min}/|V_p|},
\end{equation}
which is a function of the pulse voltage, the series resistance and the minimum voltage. Corresponding to a given pulse length $t_p$, each set of data is fitted to the curve $R(V_p)$ applying the fit parameter $|V(t_{p})|_{\rm min}$ and ${R_{\rm S}}$ by minimizing the least mean square error. As depicted in Figure~{\ref{fig_3d}}, the fitted series resistances are (averaged for pulse lengths $10^{-8}$\,s $\le t_{p} \le 10^{1}$\,s) close to the value $R_{\rm S} \approx 160$\,$\Omega$. The resulting $R(V_p)$-behavior is represented by the top edge of the colored areas in Figure~\ref{fig_3d} and match the experimental data well. In addition, it should be noticed that the estimated values of the miminum voltage for 10 and 100\,ns pulses of Figure~\ref{ps} are in good agreement with the fit parameter $|
V(t_{p})|_{\rm min}$ of Figure~{\ref{fig_3d}}.

\subsection{Switching kinetics}

\begin{figure}
\includegraphics[width=\textwidth]{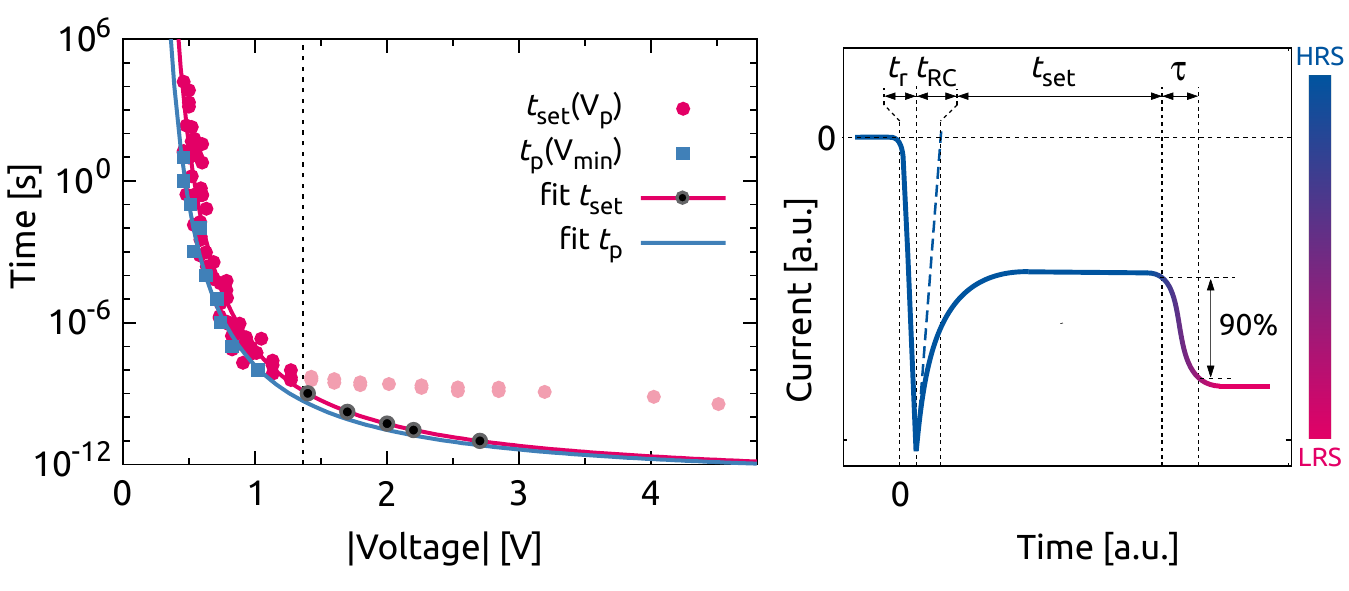}
\caption{\label{fig_5_kin} (a) Non-linear switching kinetics by the means of a $t_{\rm{set}}(V_p)$ plot over approximately 15 decades at the time scale (dark red and light red circles), and corresponding fit to Eq.\,(\ref{t-V}) for $V<1.4$\,V (red line). The 
particular fitted data pairs (black circles) are taken for the simulation in Figure~\ref{fig_6_sim}.  Furthermore, $t_{p}(V_{\rm min})$ is plotted for each $t_p$ of Figure~\ref{fig_3d} (blue squares) and fitted according to Eq.\,(\ref{t-V}) (blue line). (b) Schematic current response on a voltage step at $t = 0$ with rise time of the voltage source $t_{\rm r}$ as well as charging time $t_{\rm RC}$, set time $t_{\rm set}$, and transition time $\tau$ of the resistive cell indicating the switching event from HRS to LRS.} 
\end{figure}

By the combination of the results of different time regimes, the strong dependence of the set switching time on the pulse amplitude can be illustrated over 15 orders of magnitude (Figure~\ref{fig_5_kin}\,a). Each red colored data point represents a resistive switching event from the HRS to a state of higher conductance, which may be either the LRS or one of the IRS. To the best of the authors' knowledge it is the first time that such a high dynamic range of the switching kinetics including the picosecond regime is presented.

In the voltage range $|V| < 1.4$\,V (Figure~\ref{fig_5_kin}\,a), the experimental data show a very strong nonlinearity following the empirical relation
\begin{equation}
\label{t-V}
t_{\rm set} = t_0 \displaystyle \exp \left(\frac{\kappa}{|V_p|-V_0}\right)
\end{equation} 
with the fit parameters $\kappa=11.2$\,V and $V_0=0.162$\,V. The parameter $t_0 = 1.19\times10^{-13}$\,s is equivalent to a wavenumber $\tilde{\nu}=280\,{\rm cm}^{-1}$ for amorphous Ta$_2$O$_5$, which was found for the deformation modes of the Ta$-$O$-$Ta and Ta$\equiv$O bonds by infrared absorption spectroscopy \cite{Ono2000005}. It seems to be  that the phonon vibrations in Ta$_2$O$_5$ thin films limit the set time. \cite{Menzel2018004}

From the phyiscal point of view, the switching speed is limited by the slowest process involved. This could be e.\,g.\ the drift of oxygen vacancies or the oxygen exchange process with the electrodes \cite{Menzel2015001}. The switching speed is generally influenced by the strength of the applied electric field as well as by Joule heating enhancing the local temperature. Menzel et al.\ has shown that a temperature-accelerated drift of oxygen vacancies fits to the kinetics of a VCM cell \cite{Menzel2011001}. Actually, the Joule heating effect is playing the most significant role \cite{Menzel2015001, Witzleben2017001}. For all of these procesesses and modes of acceleration, it was theoretically derived that finally the phonon frequency will limit the switching speed \cite{Menzel2018004}. Thus, our empirical relation ({\ref{t-V}}) is consistent with the theoretical analysis. 

At higher voltage magnitudes above 1.4\,V, which correspond to shorter pulses, the measured behavior in Figure~{\ref{fig_5_kin}}\,a deviates from the expected one and the course of data points flattens towards slower switching times with increasing pulse amplitude. This is due to the fact that the voltage pulse at the device is not ideal anymore as it is, see e.\,g.\ Figure~\ref{ps}\,a, and can be qualitatively interpreted in two ways: (i) If the voltage is not fully applied over the nominal pulse length, Joule heating is reduced leading to a lower temperature, the temperature-driven ion movement is slowed down, and $t_{\rm set}$ increases. (ii) If the pulse length is too short for fully charging the cell, the ``real'' voltage over the device is lower than the nominal applied voltage, see also \cite {Lu2015001}. Therefore, the measured data pairs ($|V_p|,t_{\rm set}$) represent an upper limit of the set time at a given pulse height and should be corrected downwards as well leftwards for $|V| > 1.4$\,V. In that case, they are approaching the fitted relation of Eq.\,(\ref{t-V}). An improvement of the measurement accuracy could be realized by the means of $RC$ reduction by decreasing the cell capacitance area, however, such an approach may run into a more pronounced impedance mismatch causing a stronger damping of the transmitted signal and a worse temporal resolution.

The blue colored data pairs $t_{p}(V_{\rm min})$ in Figure~\ref{fig_5_kin}\,a illustrate the relation between pulse width and minimum voltage taken from Figure~\ref{fig_3d}. In fact, these data points behave similarly to $t_{\rm{set}}(V_p)$ and can be fitted in a similar way via Eq.\,(\ref{t-V}) with the parameter set $t_0 = 1.10\times10^{-13}$\,s, $\kappa=10.3$\,V, and $V_0=0.124$\,V. The resulting curve lies slightly below the switching kinetics data. For a given $V_{\rm min}$, the corresponding $t_{p}$ represents the moment, at which the switching does not occur anymore. Otherwise, for a given $t_{p}$, the corresponding $V_{\rm min}$ marks the voltage at which the transition halts. This analysis reveals the link between programmable resistance states and the intrinsic switching kinetics of the ReRAM cell.

\subsection{Simulation of transient currents}
\begin{figure}
\includegraphics[width=1.\textwidth]{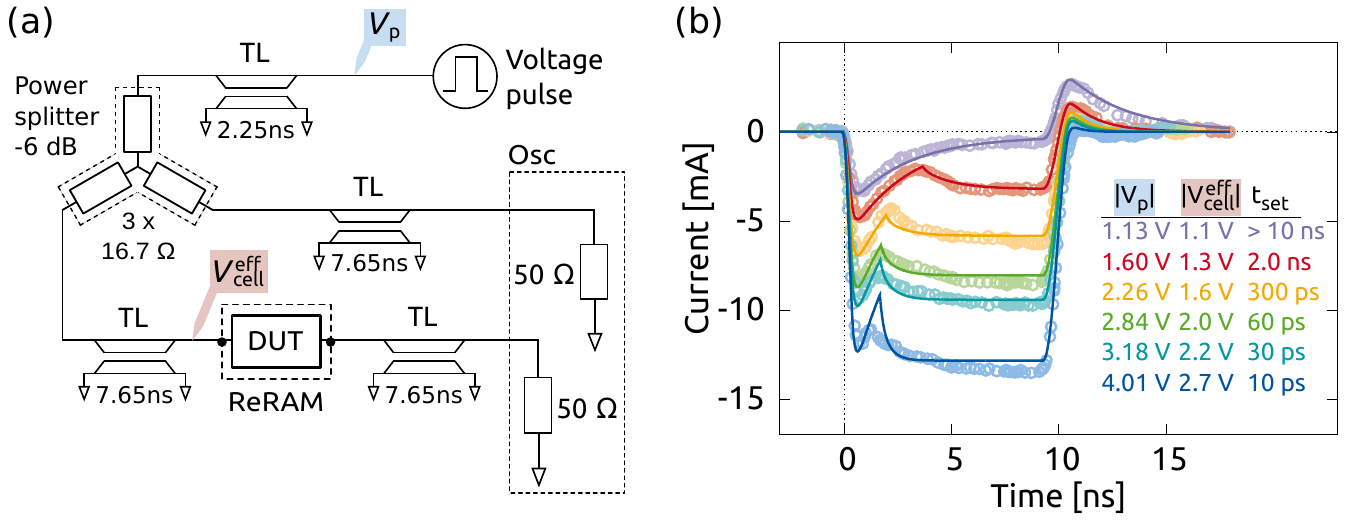}
\caption{\label{fig_6_sim} (a) Equivalent circuit of an integrated ReRAM cell (DUT) containing the relevant resistances and capacitances as described in Figure~\ref{fig_1_scheme}\,c, see also \cite{Torrezan2011001}. (b) Measured and simulated current response of memristive cell ($A=300\,\mu$m$^2$) on 10 ns pulses at different nominal voltage amplitudes $V_p$. The simulated voltage $V^{\rm eff}_{\rm cell}$ is the voltage level when the signal reaches the DUT. The applied set times $t_{set}$ result from the selected dark gray data points $t_{\rm{set}}(V_p)$ of Figure~\ref{fig_5_kin}\,a at which the voltage $V^{\rm eff}_{\rm cell}$ is taken.}
\end{figure}

To verify the suggested behavior for ultra-short pulses even at voltages $|V_p|>1.4$\,V, an empirical SPICE model simulating the resistive switching event is used. The motivation is primarily to demonstrate that the measured current responses fit  well to set times of tens or hundreds of picoseconds which result from Eq.\,(\ref{t-V}) and can be much shorter than the measured inflection points in Figure~\ref{ps}. The equivalent circuit shown in Figure~\ref{fig_6_sim}\,a comprises the series resistance $R_{\rm{S}}$, the capacitances between the CPW electrodes and the ground planes $C_{\rm{S}}$, the capacitance of the ReRAM cell $C_{\rm{cell}}$ and the time-dependent cell resistance $R_{\rm{cell}}(t)$ as described in Figure~\ref{fig_1_scheme}\,c. The time-invariant capacitances can be easily extracted from impedance measurements at 1\,MHz and are found for a cell area $A=300$\,$\mu$m$^2$ as $C_{\rm{S}} = 4.6$\,pF and $C_{\rm{cell}} = 10.6$\, pF. Taking into account the fit parameter of Figure~\ref{fig_3d}, $R_{\rm{S}}$ is 167\,$\Omega$ for the series of 10\,ns pulses.

The resistance $R_{\rm{cell}}(t)$ emulates the measured switching process, whose typical set transient behavior for long set times is shown in Figure~\ref{fig_5_kin}\,b. After charging the cell with the characteristic rise time of the voltage source $t_{\rm r}$ and the fading of the capacitive current with $t_{RC}$, the current remains initially constant in the HRS before it increases in a transition time $\tau$. Following the definition from previous publications \cite{Nishi2013002}, the switching time $t_{\rm set}$ is given as the interval between the moment the cell is charged up to 63\,\% and the onset of the current rise. For moderate pulse voltages with lengths larger than 100\,ns, $\tau$ is in the order of nanoseconds or hundreds of them depending on the applied voltage \cite{Fleck2016002, Mickel2013001}. The transition time describes the current runaway, i.\,e.\ the resistance reduction, in the moment of switching and is defined as the period from the current rise to the reaching of the 90\,\% level of the final value, which may be the LRS or one of the IRS \cite{Nishi2013001, Fleck2014002}.
 
As already mentioned, in case of ultra-short pulses, a clear identification of the characteristic times is no longer possible. The transition starts before the cell is fully charged or even during the rising edge of the voltage pulse. Therefore, the simulation is performed under the simplification of an abrupt transition ($\tau=0.1$\,ps). Figure~\ref{fig_6_sim}\,b shows a very good agreement between measured and simulated current responses of a tantalum oxide cell without any specific fitting parameters. The curves are taken for different nominal output voltages of the signal source $V_p$. The given effective voltage at the cell $V^{\rm eff}_{\rm cell}$ results from the SPICE simulation and causes the set time $t_{\rm set}(V^{\rm eff}_{\rm cell})$ calculated by Eq.\,(\ref{t-V}). It is obvious that even switching times of 10\,ps do not contradict the observed behavior. A better fit may be elaborated by applying a more accurate transition behavior.

\subsection{Integration aspects}
According to Eq.\,(\ref{Prog_R}) and assuming an invariant internal series resistance, the programmed resistance at a specific pulse width is determined by the applied voltage and the minimum voltage. Pulse width and minimum voltage, however, are not independent of each other due to the switching kinetics. If the kinetics is strongly non-linear as it is indicated by the steep slope in the log($t$)-$V$-diagram of Figure~\ref{fig_5_kin}\,a for $|V|_{\rm min} < 1$\,V and $t_{p}>100$\,ns, the minimum voltage is almost constant for all $t_{p}$ and the programmed resistance predominantly depends on the pulse voltage amplitude. For a weak non-linearity, i.\,e.\ a flat slope $d\log(t)/dV$, an additional dependence of R on the time scale is present because of the sensitivity of $|V|_{\rm min}$ to $t_{p}$. From this point of view, a highly nonlinear switching kinetics will be beneficial in terms of variability, which is permanently of major interest for resistive switching cells \cite{Nishi2018001}.

As pointed out in \cite{Kim2016004,Hardtdegen2018001}, the voltage divider effect caused by an external resistance improves the variability and the device endurance. For multilevel programming, however, a slight voltage variation close to $|V|_{\rm min}$ could evoke a larger (not acceptable) resistance variability. Thus, the pulse amplitude $|V_p|$ should be sufficient higher than $|V|_{\rm min}$. As the series resistance is linear, the resistances programmed with different voltages lie close to each other. In a big array these different resistance states might be indistinguishable considering cell-to-cell variability \cite{Calderoni2014001}. A potential strategy to overcome this problem is the use of a nonlinear series resistance. 

Without additional elements, the multilevel programming of our devices is only feasible for the set operation. As already explained above, the reset is an abrupt transition from the LRS into the HRS due to the voltage divider effect. For neuromorphic applications, however, it is desirable to program different resistances during set as well as reset operation. An suitable approach is the reduction of the voltage divider effect to emphasize the intrinsic gradual reset transition, e.\,g.\ by introducing a selector element with an asymmetric $I$-$V$ characteristics \cite{Lee2014003,Jang2018001}. For the set mode, this selector should limit the current and thus define the programmed resistance, and for the reset mode, the selector should be highly conducting that the applied voltage would drop completely over the actual resistively switching element and the intrinsic gradual reset transition appears. In this way, multilevel programming capabilities could be achieved for both voltage polarities.

\section*{Conclusion}

In this work, the multilevel resistive switching of Ta$_2$O$_5$ cells at pulse lengths down to 250\,ps was presented. For nanosecond pulses the monitoring of transient currents enables us to resolve the set switching event, and to find a clear dependence between the applied voltage and the resulting switching time. In combination with long pulse experiments, the non-linearity of the switching kinetics over 15\,orders of magnitudes was demonstrated. This behavior implies the overcome of the voltage time dilemma, which is essential for the use of any resistive two-terminal devices. A SPICE model was used to confirm that an intrinsic switching time, i.\,e.\ without any parasitic effects, of 10\,ps at 3\,V fits to the observed switching behavior. The multilevel capability together with the high write speed for a single bit provides the option to store multiple bits per cell in a time regime far below 100\,ps, which is significantly faster than writing times of state-of-the-art memory devices.

\section{Experimental}

{\bf Sample preparation.} ReRAM devices were fabricated - based on the work in \cite{Kim2016007} - by integrating a 5\,nm thin Ta$_2$O$_5$ film into a tapered 50\,$\Omega$ CPW structure designed for impedance matching of the high frequency coaxial coplanar probes (150\,$\mu$m pitch).  High-resistivity substrates of silicon (CrysTec GmbH, 4`` $\left<100\right>$ wafers, $\rho <$ 10\,k$\Omega$cm) with 450\,nm thermally grown SiO$_2$ were used. The bottom electrodes consisting of 5\,nm Ti (adhesion layer) and 25\,nm Pt were realized by DC-sputtering and patterned by ion beam etching. The deposition of the Ta$_2$O$_5$ was carried out via RF-sputtering from a metallic target with 2\% oxygen and subsequent structuring by reactive ion beam etching process. The top electrodes metals (5\,nm Ta and 25\,nm Pt) were fabricated by e-beam evaporation and lift-off lithography. Devices with effective cell areas $A=300\,\mu$m$^2$ and $600\,\mu$m$^2$ corresponding to the overlap of inner signal CPW line were processed on single wafers. All samples were initially electroformed in the HRS by a triangular positive voltage sweep using Keithley 2634B Source-Meter with an amplitude of $+4$\,V and 100\, $\mu$A current compliance.

\vspace{2ex}\noindent
{\bf Transient current response.}
The range from $10^{-7}$\,s to $10^2$\,s was characterized by pulse measurements performed with a Keithley 4200-SCS semiconductor characterization system with a 4225-PMU ultra-fast I/V modules and two 4225-RPM remote amplifiers. The transients were analyzed in terms of $t_{\rm set}$ as depicted in Figure~\ref{fig_5_kin}\,b, see also \cite{Fleck2014002,Fleck2015003}. The pulse amplitude was gradually reduced from -1 V down to -0.3 V and the pulse length proportionally varied from 100\,ns to 100\,s. For the measurements between 1\,s and 10$^5$\,s  a Keithley 2636A Souce-Meter was used. DC voltage was applied to the DUT (device under test) and the current was concurrently monitored. The voltage was varied from -0.6 V to -0.2 V in steps of 20 mV. After the detection of a significant current increase, 200 data points were subsequently recorded until the measurement was stopped. $t_{\rm set}$ was determined by use of the same algorithm as on the Keithley 4200-SCS setup.

\vspace{2ex}\noindent
{\bf Ultra-fast pulse measurements.}
Pulse generation in the nanosecond and picosecond regime applies different setups mainly based on the suggestions of Torrezan et al.\ \cite{Torrezan2011001}: (i)~a Picosecond Pulse Labs 2600C with a variable amplitude of $-45-+50$\,V at 0 dB attenuation generates pulses with widths in the range from 0.8 ns to 100 ns. The pulse amplitude can be attenuated in 1 dB steps from 0 to 70 dB. The output signal is divided by a power splitter into two identical pulses. The first part delivers the reference signal and the second pulse is guided through the DUT. (ii) pulses down to 250\,ps are generated by a Picosecond Pulse Labs 12050 pattern generator producing a continuous pattern of pulses with a width $t_p=78$\,ps, a height up to 750\,mV, and a repetition rate of $\tau_r=41$\,$\mu$s. It is combined with an appropriate timed RF switch for coupling of single pulses to the DUT. In order to provide a sufficient high set signal a PSPL 5868 RF-amplifier with 12\,V output at high impedance load is needed. The RF switch and the RF amplifier as well as the device capacitance limit the bandwidth of the system. Therefore, the minimal 78\,ps pulse width is widened to 100\,ps.  The output signal is fed directly to the DUT (without divison) since the signal reduction by the power splitter would be so strong that the required switching voltage would not be reached. Due to the capacitances of the DUT, the measured current response broadens to 250\,ps. 
The resistance of the device is in the HRS much greater than 50\,$\Omega$. Therefore the applied voltage of the SET pulse can be considered as double of the voltage assumed for an ideal 50\,$\Omega$ termination.
In both cases the signal from the DUT is captured with an ocilloscope Tektronix DPO 73304D, 33\,GHz, 100\,GS/s, real time oscilloscope with 50\,$\Omega$ input terminations. The presented transients show the current through the DUT, proportional to the voltage over the scope input. Subsequent $I(V)$ measurements by a Keithley 2634B were used to determine the resistance state after pulsing.

\vspace{2ex}\noindent
{\bf SPICE simulation.}
In order to approximate the measured transient currents, SPICE modeling was performed using the setup of Figure~\ref{fig_6_sim}\,a including the lumped elements of the ReRAM device (Figure~\ref{fig_1_scheme}\,c). The voltage pulses delivered by the PSPL 2600C are implemented as a combination of rectangular and sine sources (phase shifted sine half periods - edges, rectangular pulse - plateau), which match well with the real pulse waveform. The coaxial cables are assumed as ideal lossless 50\,$\Omega$ transmission lines with time delay of 4.5\,ns per 1\,m length. The power splitter is implemented by its resistor circuit. The abrupt resistance change from the HRS to the LRS is modelled with a transition time of 0.1\,ps.

\begin{acknowledgement}
This work was supported by the Deutsche Forschungsgemeinschaft under Grant SFB 917 ``Nanoswitches'', Project B1.
\end{acknowledgement}

\begin{suppinfo}

\begin{itemize}
  \item suppl\_info\_spice\_ecd1.pdf: Detailed equivalent circuit design including the lumped elements of the ReRAM device and the chosen parameters. The switching event is implemented by a further voltage source (rise time of 0.1\,ps) which controls the transition from the HRS to the LRS. 
  \item suppl\_info\_spice\_result.pdf: Corresponding behavior of the voltage at the DUT, the  simulated time-dependent current and the measured current behavior for $V_{p}=-1.6$\,V.
\end{itemize}

\end{suppinfo}

\bibliography{arxiv_paper_lit}
\end{document}